\documentstyle[12pt,a4wide]{article}
\begin{document}
\def\be{\begin{equation}}
\def\ee{\end{equation}}
\def\bea{\begin{eqnarray}}
\def\eea{\end{eqnarray}}
\def\coup{K}
\def\G{G}
\renewcommand{\baselinestretch}{1.}
\baselineskip 4ex
\begin{titlepage}
\begin{flushright}
KL-TH-94/8\par
CERN-TH.7290/94
\end{flushright}
\vspace{3ex}
\begin{center}
{\LARGE \bf
\centerline{ The XY Model and the }
\centerline{ Three-State Antiferromagnetic Potts Model}
\centerline{ in Three Dimensions:}
\centerline{ Critical Properties from}
\centerline{ Fluctuating Boundary Conditions}
}
\vspace{0.65 cm}
{\large    Aloysius P. Gottlob}

\vspace{0.17 cm}
{\it    Universit\"at Kaiserslautern, D-67653 Kaiserslautern, Germany}
\vspace{0.30 cm}

and

\vspace{0.30 cm}
{\large     Martin Hasenbusch\footnote{After 30. September 1994: DAMPT,
Silver Street, Cambridge, CB3 9EW, England}
}

\vspace{0.17 cm}
{\it    CERN, Theory Division, CH-1211 Gen\`{e}ve 23, Switzerland}
\end{center}
\setcounter{page}{0}
\thispagestyle{empty}
\begin{abstract}\normalsize
We present the results of a Monte Carlo
study of the three-dimensional $XY$ model and the three-dimensional
antiferromagnetic
three-state Potts model.
In both cases we compute the difference in the free energies of
a system with periodic and a system with antiperiodic boundary conditions
in the neighbourhood of the critical coupling.
 From the finite-size scaling behaviour
of this quantity we extract values for the critical
temperature and the critical exponent $\nu$ that are compatible with recent
high statistics Monte Carlo studies of the models. The results for the
free energy difference at the critical temperature and for the exponent
$\nu$ confirm that both models belong to the same universality class.
\end{abstract}
\nopagebreak
\nopagebreak
\begin{flushleft}
\vspace{0.5 cm}
KL-TH-94/8\newline
CERN-TH.7290/94\\
June 1994
\end{flushleft}
\vspace{1ex}
\end{titlepage}

\newpage
\section{Introduction}

Ueno et al. \cite{ueno} pointed out
 that the differences in the free energy  $\Delta F$
of systems
with different boundary
conditions, such as periodic and antiperiodic boundary conditions,
 might be a powerful alternative to the
 fourth-order cumulant \cite{Binder}
in the study of critical phenomena.
For the Ising model,  antiperiodic boundary conditions
 force an interface into the
system, and $\Delta F$  can be interpreted as interface free
energy. In the case of $O(N)$-invariant vector  models with $N \ge 2$,
such as the $XY$ model ($N=2$) and the classical Heisenberg model ($N=3$),
 however,
the continuous symmetry of the model prevents the
creation of a sharp interface and
 $\Delta F$  becomes rather a measure for the
helicity-modulus.

  Ueno et al. \cite{ueno} give, based on previous results
\cite{reviews},  the  scaling relation
\be
 \Delta F =  f(t L^{1/\nu}) ,
\ee
where $t=(T-T_c)/T_c$ is the reduced temperature,  $L$ the linear
extension of the lattice, and the reduced free energy $F$ is given by
$F=-\ln Z$, where $Z$ is the partition function of the system.
 It is important to note that the above relation
requires that  all directions of the lattice scale with $L$.
It follows that the crossings of $\Delta F$, plotted as
a function of
the temperature for different $L$, provide estimates for the critical
 temperature.
Furthermore the energy difference $\Delta E$,
 which is the derivative of
$\Delta F$ with respect to the inverse
temperature,
scales as
\be
 \Delta E \propto L^{1/\nu} \label{enerskal},
\ee
where $\nu$ is the critical exponent of the correlation length $\xi$.

The drawback of the method outlined above is that,
in general, it is hard to obtain  free energies   from Monte Carlo
simulations.
 The standard approach is to measure $\Delta E$ at a large number of
temperatures and perform a numerical integration starting from $T=0$
or $ T=\infty $, where the free energy is known, up to the temperature
 in question.
In \cite{Habu1,Habu2} however, one of the authors
presented a version of the cluster
algorithm \cite{Wang1,Wolff} that gives direct access to the  interface
free energy of Ising systems ($N=1$). It was demonstrated that the crossings of
$\Delta F$
converge even faster than the crossings of the fourth-order cumulant in
the case of the 3D Ising model on a simple cubic lattice.

In the present paper we show how the algorithm of refs. \cite{Habu1,Habu2}
 can be
generalized to $O(N)$-invariant vector models with $N > 1$
and apply it to the 3D $XY$ model on a simple cubic lattice.

The $\lambda$-transition of helium
from the fluid He-I phase to the superfluid He-II phase at low temperature is
supposed to share the 3D $XY$ universality class.
It is the experimentally best studied second-order phase transition.
The superfluid density corresponds to the helicity modulus of the $XY$ model
\cite{fisher}.
The quoted error bars of the measured value
  $\nu=0.6705(6)$ \cite{ahlers}
are smaller than that of the theoretical predictions for the
3D $XY$ universality class.

Banavar et al. \cite{banavar}
conjectured that the 3D antiferromagnetic (AF) three-state Potts model
belongs to the same universality class as the 3D $XY$ model.
Ueno et al. \cite{ueno} implemented ``favourable" and ``unfavourable" boundary
conditions for the
3D AF 3-state Potts  model.  They found that the corresponding $\Delta F$
is incompatible with that found for the 3D $XY$ model. They also obtained
an estimate for the critical exponent of the correlation length
$\nu =0.58(1)$ \cite{ueno},
which is not consistent with the exponent
$\nu=0.669(2)$  \cite{guillou} of  the
3D 2-component $(\phi^2)^2$-theory.
 This
result has to be compared  with recent high-precision studies of the
 3D AF 3-state Potts  model \cite{Wang2,WePott}, where the $XY$ exponents and
critical amplitudes where recovered to high accuracy.
To clarify this point we discuss how antiperiodic boundary conditions
can be implemented for the
3D AF 3-state Potts model. Our numerical findings
are then compared  with the 3D $XY$ results.

\section{$O(N)$ models with fluctuating boundary conditions}
 We consider a simple cubic lattice with extension $L$ in all directions.
 The uppermost layer of the lattice is regarded as the lower neighbour
 plane of the lower-most plane. An analogous identification is done for
 the other two lattice directions.
The $O(N)$  model is defined by the classical Hamiltonian
\begin{equation}
 H(\vec{s},bc) = - \sum_{<ij>} J_{<ij>} \vec{s}_i\cdot\vec{s}_j\;,
\end{equation}
where $\vec{s}_i$ are unit-vectors with $N$ components.
 When periodic $(p)$ boundary
 conditions $(bc)$ are employed, then $J_{<ij>}=1$
for all nearest-neighbour pairs. When  antiperiodic $(ap)$ boundary
conditions are employed, then
           $J_{<ij>}=-1$ for bonds $<ij>$ connecting the lower-most and
 uppermost plane of the
 lattice, while all other nearest-neighbour pairs keep $J_{<ij>} = 1$.
The free energy difference is now given by
\begin{equation}
 \Delta F = F_{ap}-F_{p} = - \ln \frac{Z_{ap}}{Z_p},
\end{equation}
 where $Z_{ap}$ and $Z_p$ are the partition functions with
 antiperiodic and periodic boundary conditions respectively.

 In order to obtain the ratio of partition functions
 $ Z_{ap} / Z_{p} $
  we consider a system that allows both periodic and antiperiodic
boundary conditions. The partition function of this system is given by
\begin{equation}
 Z = \sum_{bc}   \prod_{i\in\Lambda}\int_{S_{N-1}}\!\!ds_i
 \exp(-K H(\vec{s} , bc)) \, ,
\end{equation}
where $K$ is the inverse temperature.
 The fraction of configurations with antiperiodic boundary conditions
 is given by the ratio  $Z_{ap} / Z$ \, ,
\begin{eqnarray}
  \frac{Z_{ap}}{Z} &=& \frac{\prod_{i\in\Lambda}\int_{S_{N-1}}\!\!ds_i
 \exp(-K H(\vec{s},ap))} {Z}\, ,
 \nonumber \\
 &=& \frac{\sum_{bc}  \prod_{i\in\Lambda}\int_{S_{N-1}}\!\!ds_i
  \exp(-K H(\vec{s} , bc))
  \delta_{bc,ap}} {Z}\, ,  \nonumber\\ &=& \langle\delta_{bc,ap}\rangle \, ,
\end{eqnarray}
where $\delta_{bc,ap}=1$ for antiperiodic boundary conditions and
$\delta_{bc,ap}=0$ for periodic boundary conditions.
  An analogous result can be found for periodic boundary conditions.
 Now we can express the ratio $ Z_{ap} / Z_{p} $ as a ratio of
 observables in this system,
\begin{equation}
 \frac{Z_{ap}}{Z_{p}} = \frac{ Z_{ap}/Z}
                               { Z_{p}/Z}
                         =\frac{\langle\delta_{bc,ap}\rangle}
                               {\langle\delta_{bc,p}\rangle} \,
\end{equation}
which is hence accessible in a single  Monte Carlo simulation.

\section{Boundary flip algorithm for $O(N)$ models}

We shall now describe an efficient algorithm to update the system explained
 above, where the
type of boundary condition is a random variable \cite{Habu1,Habu2}.

The algorithm is based on a standard cluster algorithm
\cite{Wang1,Wolff}. For the Ising model
it can be explained as follows.
First the bonds are deleted with the standard probability
 \begin{equation}
  p_d = \exp(- K  (|s_i s_j| + J_{<ij>} s_i s_j)).
 \end{equation}
or else frozen.
After deleting or freezing the bonds of the system
one searches for an interface of deleted bonds that completely cuts
the lattice in the $z$-direction. If there is such an interface, the spins
between the bottom of the system and this interface and the sign of
the coupling $J_{<ij>}$ connecting top and bottom are flipped
simultaneously.
This is a valid update, since the bonds in the interface are deleted and
the value of $J_{<ij>} s_i s_j$, for $i$ in the lowermost and $j$ in
the uppermost plane, is not changed when we alter the sign of
$J_{<i,j>}$ and $s_i$.

In order to apply this algorithm to $O(N)$ models, each
component of the spin must be considered
 as an embedded Ising variable. In the delete
probability, we just have to replace the Ising spins by a given component of
the $O(N)$ spin.

 Note that these embedded Ising models do not couple with each other.
 The above boundary flips can be done independently for any component.

 The simplest approach would be to simulate an ensemble
 that contains
 also configurations, with different boundary conditions for the different
 components.
 However, we
 avoided these configurations with mixed boundary conditions. We only
 allowed a flip of the boundary condition,
 when it could be done for all components simultaneously.

 In our simulations we alternate this boundary flip update with
 standard single-cluster updates \cite{Wolff}.

\section{The antiferromagnetic three-state Potts model
 and antiperiodic boundary
           conditions}

The three-state AF Potts model in three dimensions is defined by the
partition function
   \begin{equation}
   Z = \prod_{l\in\Lambda}\sum_{\sigma_l=1}^{3}
   \exp\left(-\coup \sum_{\langle i,j\rangle}
   \delta_{\sigma_i, \sigma_j}\right)\; ,
   \label{a}
   \end{equation}
where the summation is
taken over all nearest-neighbour pairs of sites $i$ and $j$  on a
simple cubic lattice $\Lambda$, and
$\coup=|J|/k_BT$ is
the reduced inverse temperature.

One has to note that a change of the boundary interaction to a negative sign
is incompatible with the symmetries of the classical
Hamiltonian. The change of the sign
of $J$ from minus to plus would mean that there is only one favourable
value of the neighbouring spin instead of two.
Hence changes in the free
energy would also arise from a local distortion of the system.
However, when one adds or removes one layer from the lattice, so that
the extension in one direction, measured in units of lattice spacings,
becomes an odd number, one obtains the global frustration
we are aiming at. Hence we define $\Delta E$ of an $L^3 $ system by
  \be
   \Delta E(L,L,L) = \frac{1}{2} [E(L,L,L+1)+E(L,L,L-1)] - E(L,L,L),
  \ee
where the energy $E$ of the model is given by
  \be
  E =   \sum_{\langle i,j \rangle}
       \delta_{\sigma_i, \sigma_j} .
  \ee

We were not able to find an efficient algorithm that  adds or 
removes a layer of sites from the lattice.
Hence we had to rely on the standard integration
method to obtain the corresponding $\Delta F$
for the Potts model, as opposed to the $XY$ model.

\section{Numerical results}
\subsection{The 3D $XY$ model}

On lattices of size $L =  4,8,16,32 $ and 64, we performed simulations
at $K_0=0.45420$, which is the estimate for the critical coupling
obtained in ref. \cite{WeXY}.
 As explained above, we performed single cluster updates
\cite{Wolff} in addition to the boundary updates. We have
chosen the  number $N_0$
of the single
cluster updates per boundary update such that $N_0$ times the
average cluster volume is approximately equal to the lattice volume.
We performed a measurement after each boundary update. The number
of measurements was $100\;000$ for all lattice sizes.

First we determined the critical coupling $K_c$ using the crossings of
$Z_{ap}/Z_p$.
For the extrapolation of $\langle\delta_{bc,ap}\rangle$ and
$\langle\delta_{bc,p}\rangle$ to couplings
$K$ other than the simulation coupling $K_0$,
we used the reweighting formula \cite{swendferr}
  \begin{equation}
   \langle \delta_{bc,x} \rangle (K) =
     \frac{\sum_i \delta_{bc(i),x} \exp((-K+K_0) H_i)}
                     {\sum_i  \exp((-K+K_0) H_i)} ,
  \label{reweight}
  \end{equation}
where $i$ labels the configurations generated according to the Boltzmann weight
at $K_0$, $bc(i)$ denotes the boundary condition of the $i^{th}$ configuration,
and $x$ must  be replaced by either $p$ or $ap$.
We computed the statistical errors from Jackknife binning
\cite{siam} applied to the ratio
$\langle\delta_{bc,ap}\rangle/\langle\delta_{bc,p}\rangle$.
The extrapolation gives  good
results only 
within a small neighbourhood of the simulation coupling $K_0$. This
range shrinks with increasing volume of the lattice.
However,
fig. 1  shows that in a sufficiently large neighbourhood
of the crossings of $Z_{ap}/Z_p$
the extrapolation performs well.
The results for the crossings are $K =$ 0.45439(22), 0.45412(10),
0.454138(31), and 0.454147(14), for $L =$ 4 and 8, 8 and 16, 16 and 32,
and 32 and 64, respectively.

The convergence of the crossings of $Z_{ap}/Z_p$
towards $K_c$ is excellent. Even with the high statistical accuracy that we
reached, all crossings starting from $L=4$ and $L=8$  are compatible within
error bars.
The convergence of the crossings is governed by
  \be
  K_{cross}(L)  =  K_c \;( 1 + const.  L^{ -(\omega+1/\nu)}+\ldots),
  \label{kcross}
  \ee
where $\omega$ is the  correction to scaling exponent \cite{Binder,wegner}.
We performed a two-parameter fit with fixed $\nu=0.669$ and $\omega=0.780$
\cite{guillou}.
 Taking all crossings we obtain $K_c = 0.454142(13) $ and when
discarding the $L=4$ and 8 crossing, we get $K_c = 0.454148(15) $, where both
times the correction  term is compatible with zero.
Note that we obtained $K_c =0.45420(2)$ \cite{WeXY} (or reanalysed
$K_c = 0.45419(2)$ \cite{WePott}) from the crossing of the
fourth-order cumulant. From the scaling behaviour of the magnetic
susceptibility in the high-temperature phase we obtained $K_c = 0.45417(1)$
\cite{WeXY}. All these estimates are consistent within two standard deviations.

At the critical coupling, $Z_{ap}/Z_p$ converges
with increasing $L$ like
  \be
  \frac{Z_{ap}}{Z_p}(L)  =
   \frac{Z_{ap}}{Z_p}(\infty)  \;( 1 + const.  L^{ -\omega} \ldots) .
  \label{correction}
  \ee
In table \ref{tab1}  we give the value of $Z_{ap}/Z_p$ at our estimate of
the critical coupling.  The result is stable with increasing $L$.
Hence we take the result for $L=64$, $Z_{ap}/Z_p=0.322(8)$, as our final
estimate for the infinite volume limit. Taking the logarithm we obtain
$\Delta F = 1.13(2)$.

We extracted the critical exponent  $\nu$ of the correlation length from the
$L$ dependence
of the  energy difference $\Delta E$. The values for $\Delta E$
at the critical coupling are
given in table \ref{tab1}.
 We performed fits according to eq. (\ref{enerskal}) for the $\Delta E$ at the
 new estimate of the critical coupling and at the edges of the error bars.
The results, which are  summarized in table \ref{tab2}, are stable within
 the error bars,
 when we discard data with small $L$ from the fit.
 We take as our final result the
fit including the lattice sizes $L=16,32$, and $64$, i.e. 
$\nu = 0.679(7)$, where the error due to the uncertainty in the critical
coupling is taken into account.
Performing a similar analysis at our old estimate for the critical coupling
$K_c = 0.45419(2)$ leads to $\nu = 0.670(7)$, which is more consistent
with the accurate value $\nu = 0.669(2)$ obtained \cite{guillou} from
resummed perturbation theory.

\subsection{The 3D AF three-state Potts model}

For the 3D AF three-state Potts model we computed $\Delta F$ by the integration
method. At $K=0$ the free energy is given by
\begin{equation}
F = V \ln 3\, ,
\end{equation}
where $V$ is the number of lattice sites.
Hence
\begin{equation}
\Delta F =  \frac{1}{2} [F(L,L,L-1) + F(L,L,L+1)]- F(L,L,L) = 0
\end{equation}
at $K=0$.
For $L=4$ we measured $\Delta E$ at 83 different values of $K$, starting at
$K=0.01$ and going up in steps of $\Delta K=0.01$ until we reached $K=0.83$.
 In the large-$L$ limit, $\Delta F$ stays $0$ up to the critical
point. Therefore
we started the integration at a $K$ such that
we observed $\Delta E > 0$ within our statistical accuracy for the larger
lattices.
For $L=8$ we measured $\Delta E$ at 67 different values of $K$, starting at
$K=0.50$ and going up in steps of $\Delta K=0.005$ until we reached $K=0.83$.
For $L=16$ we measured $\Delta E$ at 52 different values of $K$, starting at
$K=0.70$ and going up in steps of $\Delta K=0.0025$ until we reached $K=0.83$.

All runs  consisted of $10\;000$  measurements. Per measurement we performed
such a number of single cluster updates
 that the lattice volume was approximately
covered by the  average cluster volume. Then we performed the integration using
the trapeze rule. The result is given in fig. 2. The curves for $L=4$ and
$L=8$
cross at $K=0.8155(18)$ and the curves for $L=8$ and $L=16$ at $K=0.8166(8)$,
which is in good agreement with  $K_c = 0.81563(3)$ \cite{WePott}.
 The values of $ \Delta F$
at $K_c = 0.81563$ are summarized in
table 3.
Our statistical accuracy degrades with increasing lattice size. Hence
we skipped the simulations of larger lattice sizes.
However,
 already for the small lattices the results at $K_c = 0.81563$ are rather
stable; systematic errors due to corrections to scaling should be small.
The result $\Delta F = 1.13(2)$ for  $L=16$ at the critical point
nicely agrees with our
 final result $\Delta F = 1.13(2)$ for the 3D $XY$ model.
One should note that for the 3D Ising model one obtains $\Delta F = 0.605(6)$
\cite{Habu2}, which is only a little more than half the $XY$ value.

At $K_c = 0.81563$ we simulated the $L \times L \times L-1$ and
$L \times L \times L+1$ lattices for sizes up to $L=64$ with
a statistics of $100\;000$ measurements. For the cubic lattices we
used the results of our  previous study \cite{WePott},
 where $200\;000$ measurements had been
 performed.  The resulting $\Delta E$ are summarized in table \ref{tab3}.
We performed fits according to 
eq. (\ref{enerskal}) for the $\Delta E$ at the
critical coupling and at the edges of the error bars.
The results are  summarized in table \ref{tab4}. The fit including all lattice
sizes gives an unacceptably large $\chi^2/degrees\; of \; freedom$,
which in the following will be denoted as $\G$. Discarding the $L=4$ data
the value of $\G$ becomes acceptable, and when discarding also the $L=8$ data
the result for $\nu $ remains stable within the error bars.
Hence we conclude that
systematic errors due to corrections to scaling are smaller than our
statistical errors.
We take $\nu=0.663(4)$, from
the fit including the lattice sizes $L=16,32$ and $64$, as
our final result, where the error due to the uncertainty in the critical
coupling is taken into account.

\section{Conclusions}

In the present work we have shown how the boundary algorithm of refs.
\cite{Habu1,Habu2} can be applied to $O(N)$ models with $N > 1$.
 We demonstrated, in the case of the 3D $XY$ model, that its critical
properties can be nicely extracted from the ratio of the
partition functions $Z_{ap}/Z_p$. The accuracy of the
results for the critical coupling
and the critical exponent of the correlation length $\nu$  are compatible
with that obtained from the fourth-order cumulant.

We showed how antiperiodic boundary conditions can be implemented for the
3D AF Potts model. The value of the free energy difference
 $\Delta F = F_{ap}-F_{p}$
at the critical coupling is in good agreement with that found for the
3D $XY$ model.
  The value $\nu = 0.663(4)$ obtained from the scaling behaviour of
 the energy difference $\Delta E$ at the critical coupling is as
accurate as  our previous estimate, which we obtained from the slope of the
fourth-order cumulant. We conclude that this strongly supports the fact that
the 3D AF 3-state Potts model and the 3D $XY$ model belong to the same
universality class.
The confirmation of the conjecture by Banavar et al.  also has practical
implications. The 3D AF 3-state Potts model is simpler to simulate than
the 3D $XY$ model. The application of multispin-coding techniques,
which  have been used to speed up simulations of the Ising model \cite{Ito},
might also allow further improvements  of the  3D AF 3-state Potts 
results. A first attempt in this direction you can find in ref.\cite{okabe}.

For a detailed comparison with previous results, see \cite{WePott}.

\section*{Acknowledgements}

We would like to thank D.~Stauffer for many helpful suggestions.

The major part of the numerical simulations was performed on an
 IBM RISC 6000 cluster of the
Regionales Hochschulrechenzentrum Kaiserslautern (RHRK).
The simulations took about one CPU-month on
an IBM RISC 6000-590 workstation.


\newpage
%
\begin{figure}
\caption{
The ratio $Z_{ap}/Z_p$ for the 3D $XY$ model on  lattices of size
 $L=4$ up to $L=64$. The curves are obtained from simulations at $K=0.45420$
in combination with reweighting to couplings in the neighbourhood.
The dashed lines give the statistical
errors obtained by a Jackknife analysis.
\label{histou1}}
\end{figure}
%
\begin{figure}
\caption{
The free energy difference  $\Delta F$ for the 3D AF Potts model
 on  lattices of size
 $L=4, 8$, and 16. The curves are obtained from numerical integration of
$\Delta E$.
The dashed lines give the statistical
errors obtained by a Jackknife analysis.
\label{histou2}}
\end{figure}
 
\newpage


\begin{table}
\caption{Results of the ratio $Z_{ap}/Z_p$ and $\Delta E$
at the critical
coupling $K_c=0.45415(2)$. The number in the second bracket gives the
uncertainty due to the error bar of the critical coupling.}
\label{tab1}
\begin{center}
\begin{tabular}{rll}\hline\hline
$L$ & $ Z_{ap}/Z_p$ & $ \Delta E $ \\
\hline
4 &  0.3245(19)(1) & 17.64(13)(1)\\
8 & 0.3242(19)(3) &  52.18(41)(4)\\
16& 0.3234(21)(9) &  144.7(1.3)(3)\\
32& 0.3224(20)(27) & 407.0(4.6)(2.2) \\
64& 0.3216(25)(72) &  1113.(13.)(18.)\\ \hline\hline
\end{tabular}
\end{center}
\end{table}

\begin{table}
\caption{
         Estimates of the critical exponent $\nu$ obtained from the
         fit of the surface energy density following
         eq. (\protect\ref{enerskal})
         at $K_c=0.45415(2)$. \# gives the number of discarded
         data points with small $L$ and $\G$ denotes
         $\chi^2 / degrees\; of \; freedom$.
        }
         \label{tab2}
\begin{center}
\begin{tabular}{ccccccc}\hline\hline
&\multicolumn{2}{c}{ $K_c-\Delta K_c$}&
 \multicolumn{2}{c}{ $K_c$}&
 \multicolumn{2}{c}{ $K_c+\Delta K_c$}\\ \hline
\#    & $\nu$       &$\G$& $\nu$
& $\G$  & $\nu$ & $\G$  \\ \hline
 0&0.6771(45) &  1.80 &
   0.6756(44)  &  0.78 &
   0.6741(44)  &  0.77 \\

 1&0.6813(30)  &  0.96 &
   0.6783(29)  &  0.53 &
   0.6753(29)  &  0.32 \\

 2&0.6833(50)  &  1.67 &
   0.6787(49)  &  1.05 &
   0.6743(48)  &  0.57 \\ \hline\hline
\end{tabular}
\end{center}
\end{table}

\begin{table}
\caption{$\Delta F$ and $\Delta E$
for the 3D AF 3-state Potts model at $K=0.81563(3)$. The number in the second
bracket of $\Delta E$ gives the uncertainty due to the error bar of the
critical coupling.
         }
\label{tab3}
\begin{center}
\begin{tabular}{rll} \hline\hline
 $L$ &  $\Delta F$ & $\Delta E$ \\ \hline
 4   &  1.165(8) &    6.68(2)     \\
 8   &  1.157(13)&   19.10(9)(1)    \\
16   &  1.130(20)&   53.36(30)(6)   \\
32   &           &  152.92(92)(48)  \\
64   &           &  431.8(2.6)(3.7) \\ \hline\hline
\end{tabular}
\end{center}
\end{table}

\begin{table}
\caption{Results for the critical exponent $\nu$ obtained from the fit
         following eq. (\protect\ref{enerskal}). \# denotes the number of
         discarded data points with small $L$ and $\G$ denotes
         $\chi^2 / degrees$ $of \; freedom$.
}
\label{tab4}
\begin{center}
\begin{tabular}{rlclclc} \hline\hline
\phantom{$\#$}&\multicolumn{2}{c}{$K_c-\Delta K_c$}
            &\multicolumn{2}{c}{$K_c$}
            &\multicolumn{2}{c}{$K_c+\Delta K_c$}\\
 \hline
$\#$ &\phantom{$123$} $\nu$           & $\G$
   & \phantom{$123$} $\nu$         & $\G$
   & \phantom{$123$} $\nu$         & $\G$ \\ \hline
 0 & 0.8988(49)  &  2528
   & 0.9130(43)  &  3392
   & 0.8598(42)  &  1903 \\
 1 & 0.6681(19)  &  1.21
   & 0.6664(16)  &  1.92
   & 0.6650(18)  &  1.73 \\
 2 & 0.6651(32)  &  1.10
   & 0.6629(26)  &  1.04
   & 0.6605(31)  &  0.45 \\ \hline\hline
\end{tabular}
\end{center}
\end{table}
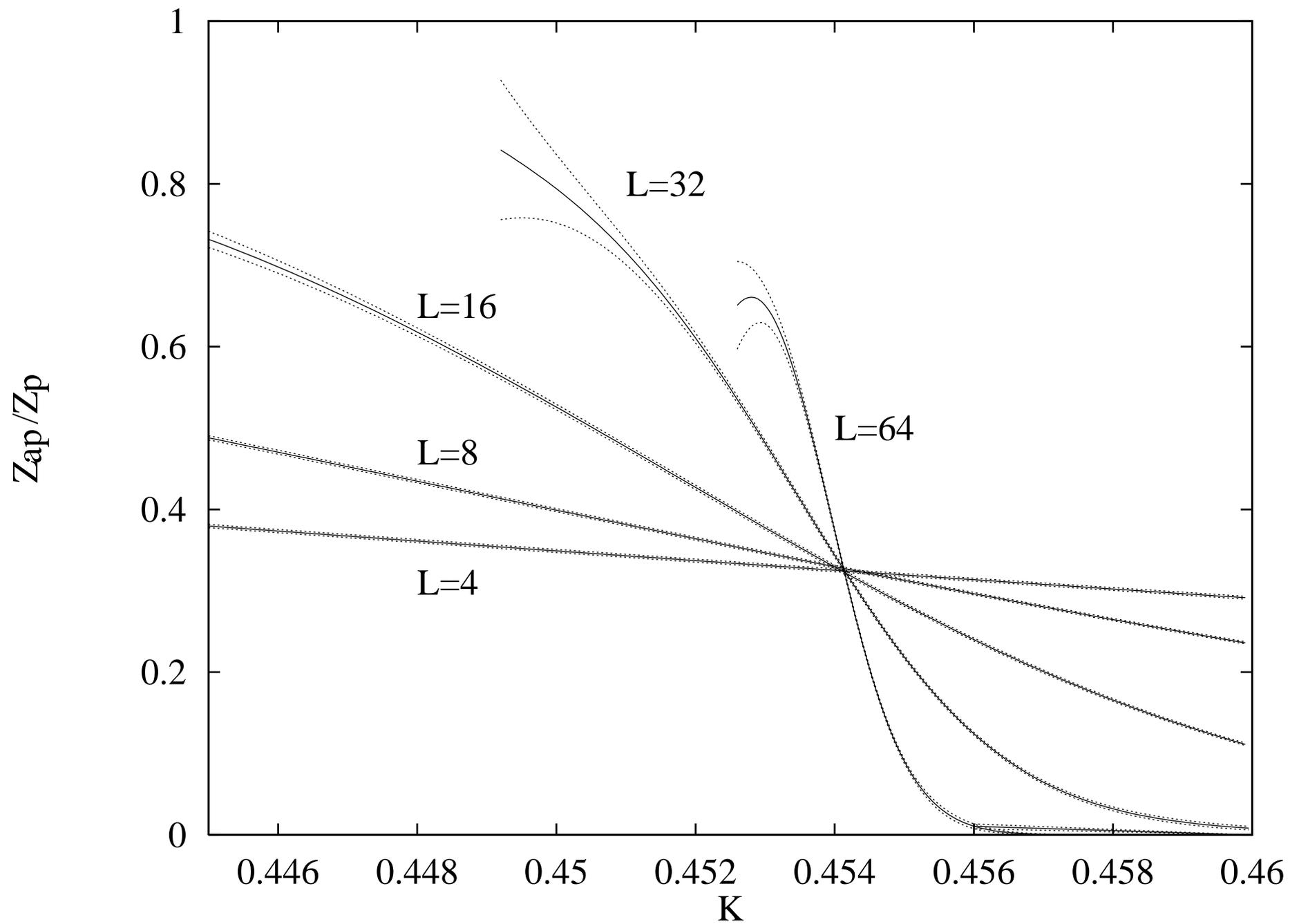
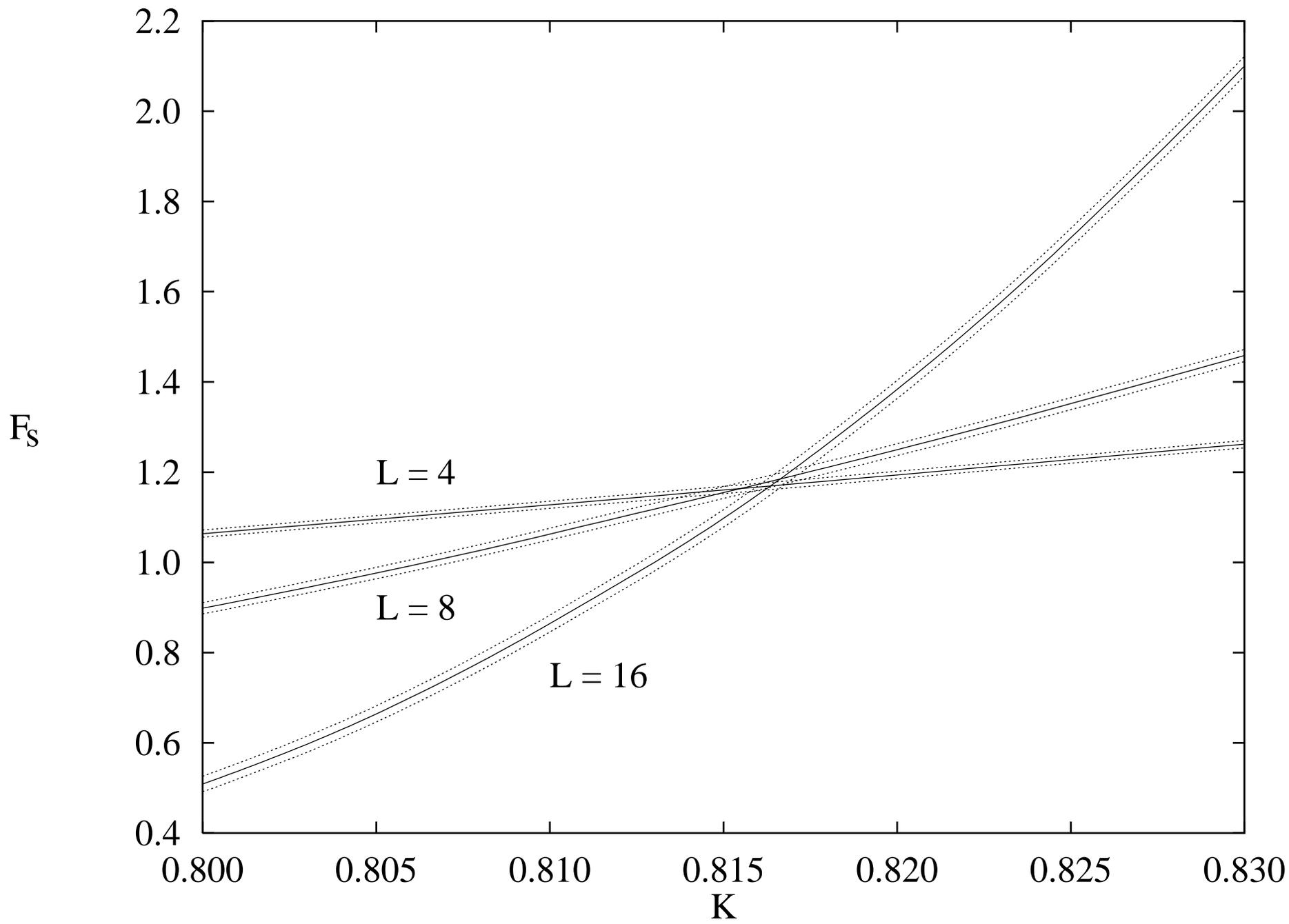
\end{document}